\documentclass[twocolumn,showpacs,preprintnumbers,amsmath,amssymb]{revtex4}

\usepackage{graphicx}
\usepackage{dcolumn}
\usepackage{bm}

\begin{document}

\def\relR{\mbox{\small \bf R}}
\def\inr{\mbox{\small \bf r}}
\def\tn{\tilde n}
\def\Rdot{\stackrel{\cdot }{\relR}}
\def\Rddot{\stackrel{\cdot \cdot }{\relR}}

\preprint{APS/123-QED}

\title{The Role of Mass Asymmetry and Shell Structure\\
in the  Evaporation Residues Production}

\author{G.~Fazio}
\author{G.~Giardina}%
 \email{giardina@nucleo.unime.it}
\author{A.~Lamberto}%
\affiliation{%
INFN, Sezione di Catania,\\
and Dipartimento di Fisica dell'Universit{\`a} di Messina, Italy
}%

\author{A.I.~Muminov}
\author{A.K.~Nasirov}%
\altaffiliation[Also at ]{Bogoliubov Laboratory of the Theoretical Physics,
JINR, Dubna, Russia, and
Institut f{\"u}r Theoretische Physik der Justus-Liebig-Universit{\"a}t,
Giessen, Germany}
\affiliation{%
Heavy Ion Physics Department, INP, Tashkent, Uzbekistan
}%

\author{U.T.~Yakhshiev}
\affiliation{%
Theoretical Physics Department, National University of Uzbekistan, Tashkent, Uzbekistan
}%

\author{Yu.Ts.~Oganessian}%
\author{A.G.~Popeko}%
\author{R.N.~Sagaidak}%
\author{A.V.~Yeremin}%
\affiliation{%
Flerov Laboratory of Nuclear Reactions, JINR, Dubna, Russia
}%

\author{S.~Hofmann}
\affiliation{
Gesellschaft f{\"u}r Schwerionenforschung mbH, Darmstadt, Germany
}%

\author{F.~Hanappe}
\affiliation{
Universit\'e Libre de Bruxelles, Bruxelles, Belgium
}%

\author{L.~Stuttg\'e}
\affiliation{
Institut de Recherches Subatomiques, Strasbourg, France
}%

\date{\today}

\begin{abstract}
The effects of the entrance channel and shell structure
of reacting nuclei on the experimental  evaporation residues
have been studied by analysing the  $^{40}$Ar + $^{176}$Hf,
$^{86}$Kr + $^{130,136}$Xe,  $^{124}$Sn + $^{92}$Zr
and $^{48}$Ca + $^{174}$Yb reactions leading to
the $^{216}$Th$^*$ and $^{222}$Th$^*$ compound nuclei.
The measured excitation function of evaporation residues
for the $^{124}$Sn + $^{92}$Zr reaction was larger
than that for the  $^{86}$Kr +  $^{130}$Xe reaction.
The experimental values of  evaporation residues in the
$^{86}$Kr + $^{136}$Xe reaction were about 500 times
larger than that in  the $^{86}$Kr + $^{130}$Xe  reaction.
These results   are explained by the initial angular
momentum dependence of the fusion excitation
functions calculated in framework of the dinuclear system concept
and by the differences in survival probabilities calculated
in framework  of advanced statistical model.
The dependencies of the fission barrier and the
$\Gamma_n / \Gamma_f$ ratio on the angular momentum of the excited
compound nucleus are taken into account.
\end{abstract}

\maketitle

\section{\label{intro}Introduction}

The study of the role of entrance channel in the formation
of evaporation residues is an actual problem to establish  the optimal
conditions for the synthesis of new superheavy elements.
Comparison of excitation functions of evaporation residue  (ER)
measured for different mass-asymmetry reactions but leading to the
same compound nucleus (CN) allows us to analyze the importance of
the entrance channel effect on the fusion-fission reaction
mechanism in collisions of massive nuclei. Often excitation
functions of evaporation residues, measured in various reactions
leading to the same compound nucleus, are different not only in
the position of the maximum but also in the value of their
maximums.

The analysis of data obtained from experiments in GSI (Darmstadt) and
Flerov Laboratory of Nuclear Reactions (Dubna)
reveals  that the maximum value of the ER
cross section for $^{40}$Ar+$^{176}$Hf  \cite{Verm84,Clerc84}
is twelve   times larger than for $^{86}$Kr + $^{130}$Xe
\cite{Ogan96} and three  times  larger than for $^{124}$Sn + $^{92}$Zr
 \cite{Sahm85}. All of these reactions lead to the same excited
 $^{216}$Th$^*$ compound nucleus. The $^{40}$Ar+$^{176}$Hf
reaction has a larger mass  asymmetry
($\eta_{A}=(A_{2}-A_{1})/(A_{1}+A_{2})$) in comparison
with the two others.
But intriguing phenomenon is that the measured
maximum value of the ER for $^{124}$Sn + $^{92}$Zr is four times
larger than for $^{86}$Kr + $^{130}$Xe, nearly at the same
$E^*$ value,  though  the mass asymmetry $(\left|\eta_A\right|=0.148)$  of
the $^{124}$Sn + $^{92}$Zr reaction is smaller
than the one of $^{86}$Kr + $^{130}$Xe (0.204).

In case of  $^{48}$Ca + $^{174}$Yb   \cite{Sagai97} and
$^{86}$Kr + $^{136}$Xe \cite{Ogan96} reactions leading to the excited
$^{222}$Th compound nucleus, the  comparison of the measured data on the cross
section of evaporation residues does not  show strongly the role of
mass asymmetry of entrance channel.

The influence of the neutron number on the measured ER was
studied in reactions with $^{86}$Kr on the $^{130}$Xe and
$^{136}$Xe targets.
The difference  between the experimental data in above mentioned
reactions shows to be connected
by the dependence of the strength of quasifission on the
entrance channel, namely on the mass asymmetry and shell
structure of colliding nuclei.

The entrance channel dependence of the
distribution of reaction strength has been studied for three
systems, namely $^{32}$S + $^{182}$W, $^{48}$Ti + $^{166}$Er, and
$^{60}$Ni + $^{154}$Sm, which all lead to the compound system
$^{214}$Th in complete fusion reactions \cite{Back96}. The
maximum contribution of complete fusion-fission process to the
fission-like cross section is estimated on the basis of expected
angle-mass correlations for such reactions. The results show a
strong entrance channel dependence as predicted by the extra-push
model.

The role of entrance channel effects was studied in
\cite{Hahn87}
where  the reactions with  $^{40}$Ar and $^{84}$Kr leading
to the same $^{200}$Po  CN were analysed. Comparison of the measured excitation
functions for the isotopes $^{200-xn}$Po  produced  in the $^{40}$Ar +
$^{160}$Dy and  $^{84}$Kr + $^{116}$Cd  reactions showed that the
(Ar, $x$n) cross sections are larger by factors of  $2\div4$
than the corresponding  (Kr,$x$n) values. In the experiment of
this group, an effect of the entrance channel on the formation and
decay of  $^{158}$Er produced in reactions with either  $^{40}$Ar
and $^{84}$Kr, as well as on the de-excitation of  compound nucleus
by neutron evaporation was not considered \cite{Hahn79}.
In reactions of massive projectile and target nuclei, the competition between
complete fusion and quasifission  appears  at the stage of CN
formation, in addition to the increase of  its fission probability.
Even in the case of mass-asymmetric collisions, an inhibition of the
fusion was recently observed
in the experiment by  Hinde and  his colleagues \cite{Berriman}.
The $^{12}$C +$^{204}$Pb, $^{19}$F + $^{197}$Au and $^{30}$Si + $^{186}$W
reactions leading to the same  $^{216}$Ra nucleus  have  been studied.
The authors stressed that there is a significant inhibition of the reduced fusion
cross section ($\tilde\sigma=\sigma/\pi\lambda\hspace*{-2.5mm}-^2)$
for reactions with  $^{19}$F  and $^{30}$Si, being $(0.64\pm0.09)$ and
$(0.57\pm0.08)$, respectively, of those for $^{12}$C.

According to the macroscopic dynamical model  (MDM) \cite{Block86}
the "extra push" energy, which is needed to
transform dinuclear system into compound nucleus, is smaller for
asymmetric reaction than for more  symmetric one leading
to the same compound nucleus.
The same effect of the mass asymmetry of the projectile-target pair
on the probability of compound nucleus formation is obtained
in models based on dinuclear system concept (DNS)
\cite{DNSV935,DNSCh96,Adam98,GiarSHE,Cher99}. In the DNS concept,
fusion  is  considered as nucleon transfer from the light fragment to
the  heavy one. From these theories it follows that in  reactions with nuclei
of  symmetric masses there is an inhibition for the fusion, and the
quasifission appears as a competing channel with complete fusion.

Quasifission reactions are binary processes
that exhibit some of the characteristics of fusion-fission events,
such as the full relaxation of the relative kinetic energy and a
considerable transfer of mass between the two fragments. The basic
difference between fusion-fission and quasifission is that
compound nucleus formation is not achieved in the latter
mechanism. Anyhow, it is difficult to establish directly
in the experiment  the origin of fusion-fission fragments.

The aim of this paper is to analyse the role of the mass asymmetry
and shell structure in fusion-fission reactions by comparison of the
difference between the experimental data  for  the $^{40}$Ar +
$^{176}$Hf \cite{Verm84,Clerc84}, $^{86}$Kr + $^{130}$Xe
\cite{Ogan96} and $^{124}$Sn + $^{92}$Zr \cite{Sahm85} reactions
leading to the $^{216}$Th$^*$ compound nucleus and $^{48}$Ca +
$^{174}$Yb \cite{Sagai97} and $^{86}$Kr + $^{136}$Xe \cite{Ogan96}
reactions leading to the $^{222}$Th$^*$ compound nucleus.

A  model based on the DNS concept \cite{DNSV935}  allows one
to estimate both contributions of fusion-fission and quasifission
processes.  It reveals the  competition between these processes
for a massive system or for a symmetric entrance channel in the case
of mid-heavy systems.
Calculations based on the DNS-concept show
\cite{GiarArAg,DNSCh96,Adam98,GiarSHE,Cher99} that entrance channel
effects are important to describe the experimental data in the
case of collisions of  massive nuclei. It allows us to estimate
the decrease of the fusion probability due to increase of the quasifission
process.
Calculations of the competition between complete fusion and
quasifission process  include the peculiarities of shell structure
and shape of colliding nuclei. That allows us to reach useful
conclusions about the
mechanism of the fusion-fission process. As an example,  the measured
fission excitation function for the $^{40}$Ar+$^{176}$Hf reaction
 obtained  from the detection of reaction
products of symmetric masses is compared with
 the calculated  fusion excitation function.
It should be stressed that those products  could be formed
not only  at  the fission of a hot CN but at
quasifission of DNS which lives long enough to
reach mass equilibration in the subsequent re-separation process.

The dynamical approach to the formation and evolution of
DNS \cite{GiarArAg} in a pair  with the advanced
statistical model  (ASM) \cite{ASM} shows a good agreement of the
calculations with the experiments in a mid-heavy (non-fissile)
region of CN \cite{GiarArAg}.

The structure of the article is as follows. Basic features of the
dynamical approach and the advanced statistical model  are
described in Section~\ref{dnscon}. In Section~\ref{dnsexp}
and \ref{130136xe}, we
compare the results of calculation with the experimental data and
discuss the  effect of entrance channel  on reaction mechanism.
Conclusions are presented in Section \ref{concl}.

\section{\label{dnscon}Evaporation residue production in the DNS concept}

According to the DNS concept, evaporation residue production  is
considered as a three stage process. The first step is overcoming
the Coulomb barrier in motion along the axis connecting nuclear
centers by nuclei at the incoming stage of collision, and
formation of nuclear composite (molecular-like so-called dinuclear
system). This stage is called capture. The second one is
transformation of the DNS into more compact compound nucleus
in competition with quasifission process. At this stage, the system
must overcome  the intrinsic barrier ($B^*_{fus}$) on the potential
energy surface during evolution on mass (charge) asymmetry axis.
For light and intermediate nuclear systems or for heavy nuclear
systems with larger mass asymmetry, this  barrier is equal to zero
and capture immediately leads to fusion. Therefore, in those
cases, the fusion cross section is calculated in the framework of
well known models \cite{Swiat812,Davie83,Frob848,Block86}.
This barrier will be discussed later.
 It should be stressed that complete fusion  is a  transfer of all
 the nucleons of the projectile
(or light nucleus)  into the target. Due to large inertia parameter
of deformation the change of nuclear shape from the initial state
 is not so  large and
the size of overlap region of nuclei is small: it is about 5-6\%
of the total volume. Therefore,  the interacting nuclei  retain
their shell structure  during interaction.

At the third stage, the hot compound nucleus  cools down by emission
of neutrons and charged particles. There is a chance of nucleus to
undergo fission at each step of the de-excitation cascade.
Therefore the evaporation residues cross section is determined by
the partial fusion cross sections and survival probabilities
of the excited compound nucleus:

\begin{equation}
\label{evapor}
\sigma_{er}(E)=\sum_{\ell=0}^{\ell_d}(2\ell+1)\sigma_{\ell}^{fus}(E)
W_{sur}(E,\ell).
\end{equation}
Here, the effects connected with the entrance channel  are included
in the partial fusion cross section $\sigma_{\ell}^{fus}(E)$, which
is defined by the product  of partial capture cross sections  and the
related fusion factor  ($ P_{CN}$)  taking into account competition between
complete fusion and quasifission processes:
\begin{eqnarray}
\label{s_fus}
\sigma_{\ell}^{fus}(E)&=&\sigma_{\ell}^{capture}(E) P_{CN}(E,\ell),\\
\sigma_l^{capture}(E)&=&\frac{\lambda^2}{4\pi}{\cal
P}_{\ell}^{capture}(E).
\end{eqnarray}
Here $\lambda$ is the de Broglie wavelength of the entrance
channel;  ${\cal P}_{\ell}^{capture}(E)$ is the capture
probability which depends on the  collision dynamics and is
determined by the number of partial   waves ($\ell_d$) leading to
capture.

\begin{figure}
\includegraphics[totalheight=11cm]{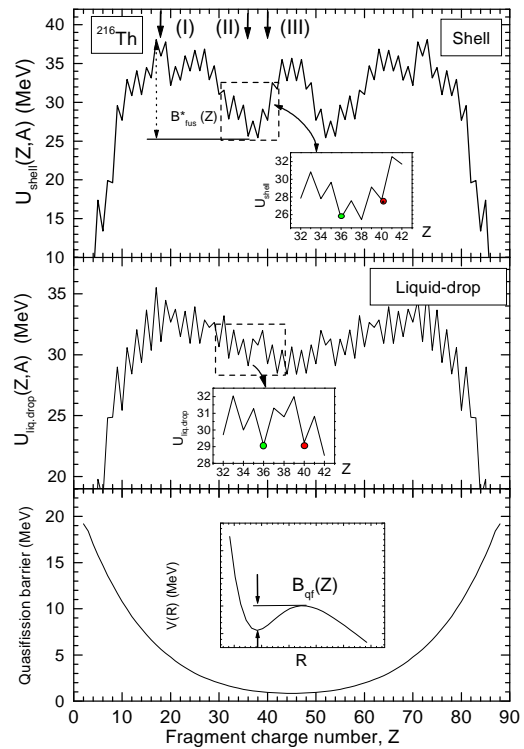}
\vspace{-0.5cm} \caption{\label{dr216}Driving potential
$U(Z,A,R_m;l=0)$ as a function of the charge number  $Z$ of a
fragment of a DNS calculated by (\ref{PES}) using binding energies
from the nuclear data in \cite{MassAW95} (top panel) and those
obtained with the liquid-drop model (middle panel). The vertical
arrows indicate the initial charge number of light nuclei in the
$^{40}$Ar+$^{176}$Hf (I) \cite{Verm84,Clerc84},
$^{86}$Kr + $^{130}$Xe (II) \cite{Ogan96} and
$^{124}$Sn + $^{92}$Zr (III) \cite{Sahm85} reactions leading to
$^{216}$Th. The intrinsic  $B^*_{fus}$ (top panel) and
quasifission  $B_{qf}$ (bottom panel) barriers are shown as a
function of the charge number of a DNS fragment.}
\end{figure}

\begin{figure}
\includegraphics[totalheight=10cm]{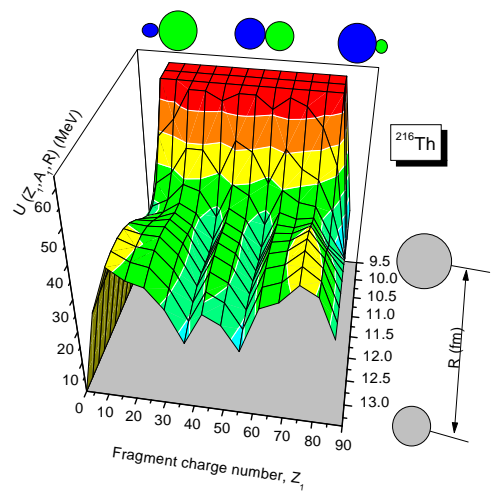}
\vspace{-3cm}
\caption{\label{pes216}Potential energy surface $U(Z,A,R;l=0)$
as a function of the distance $R$ between the centers of the nuclei
and charge number $Z$ of a fragment.}
\end{figure}

The potential energy surface is built as a function of the mass
(charge) asymmetry and relative distance between centers of nuclei
forming DNS.
$B^*_{fus}$ is determined by the difference between the
maximum value of the driving potential   $U(Z,A,R_m)$ and its
value at the point corresponding to the initial charge asymmetry
of the considered reaction (Fig.\ref{dr216}). For example, in top panel
of Fig.\ref{dr216}, $B^*_{fus}$ is shown for
the $^{86}$Kr + $^{130}$Xe reaction. If the excitation
energy of dinuclear system, $E^*_{DNS}=E_{c.m.}-V(R_m,\ell)$, is
not enough to overcome $B^*_{fus}$ then the dinuclear system
may immediately decays into two fragments or its  decay occurs
after multi-nucleon transfer from heavy fragment into light one.
Both of decays are called quasifission. So quasifission
fragments can be of different mass asymmetry.
Quasifission occurs due to  motion along relative internuclear
distance  $R$ and DNS should overcome the barrier ($B_{qf}$) defined
by the depth of well of $V(R)$ (see insert in bottom
panel of Fig. \ref{dr216}).

The $U(Z,A,R_m)$ is extracted from the potential energy surface
$U(A,Z;R,\ell)$ (\ref{PES}), which is a function of mass $A=A_1$
(or $A_2=A_{CN}-A$) and charge $Z=Z_1$ (or $Z_2=Z_{CN}-Z$)
of one of fragments forming the DNS at
the values $R_m$ of the internuclear distance corresponding to the
minimum of their nucleus-nucleus potential $V(R)$ (\ref{nucnuc});
$A_{CN}$ and $Z_{CN}$ are mass and charge of compound
nucleus, respectively.
The total  potential energy $U(A,Z,R;\ell=0)$, calculated
by this way for the $^{216}$Th CN, is presented in
Fig.\ref{pes216}. The distribution of neutrons between two
fragments for the given proton numbers  $Z$ and $Z_2$ (or ratios
$A/Z$ and $A_2/Z_2$ for both fragments) was determined
by minimizing the potential $U(A,Z;R)$ as a function of $A$
for each $Z$:
\begin{eqnarray} \label{PES}
U(A,Z;R,\ell)&=&U(A,Z,\ell,\beta_1,\alpha_1; \beta_2,\alpha_2)
\nonumber\\
&=&B_1+B_2+V(Z,\ell,\beta_1,\alpha_1; \beta_2,\alpha_2; R)
\nonumber\\
&-&(B_{CN}+V_{CN}(\ell)).
\end{eqnarray}
Here, $B_1$,  $B_2$ and $B_{CN}$ are the binding energies of the
nuclei in   DNS and of the CN, respectively, which were obtained
from \cite{MassAW95,Moeller};  $\beta_i$ are  the fragment
deformation parameters and $\alpha_i$ are the orientations
relative to the beam direction; $V_{CN}(\ell)$ is the rotational
energy of the compound nucleus. The $R_m$ is the position of this
minimum (bottom of the pocket) on the $R$ axis for a given mass
of fragment $A$. The smallest excitation energy value of the CN is
determined by the absolute maximum value of the driving potential
lying on the way to fusion ($Z=0$)  from the point corresponding
to the initial charge asymmetry (Fig.\ref{dr216}). The shapes of
the potential energy surface and
driving potential depend on the orientations of nuclei relative to
the axis connecting the centers of interacting nuclei. The
presented results were obtained by averaging the contributions of
different orientations.

\subsection{\label{capture} Capture}

The capture cross section is defined by the number of partial waves
which lead colliding nuclei to trap into the  well of
the nucleus-nucleus potential.
The number of the partial waves $\ell_d$ was obtained by   solving
the equations of motion for the relative distance and orbital
angular momentum taking into account dissipation of collective kinetic energy
\cite{GiarArAg,GiarSHE}:
\begin{eqnarray}
\label{maineq}
&&\mu(R(t))\Rddot + \gamma_{R}(R(t))\Rdot(t)=
-\frac {\partial V(R(t))}{\partial R},\\
&&\frac{dL}{dt}=\gamma_{\theta}(R(t))\left(\dot{\theta} R_{eff}^2
-\dot{\theta_1} R_{1eff}^2
-\dot{\theta_2} R_{2eff}^2\right)\,,
\end{eqnarray}
where $R(t)$ is the relative motion coordinate; $\Rdot(t)$ is
the corresponding velocity; $\dot\theta$, $\dot\theta_1$ and
$\dot\theta_2$ are  angular velocities  of the DNS and
its  fragments, respectively;  $\gamma_{R}$  and $ \gamma_{\theta}$
are the friction coefficients for the relative motion along $R$ and
the tangential motion when two nuclei roll on each other's surfaces,
respectively;
$V(R)$ is the nucleus-nucleus potential; $\mu(R(t))$ is the reduced
mass of the system:
\begin{equation}
\label{massredu}
\mu(R)= m_0 A_T A_P/A_{tot}\\
\end{equation}
where $A_{tot}=A_T+A_P; m_0$ is the nucleon mass; $A_T$ and $A_P$ are
mass numbers of the target- and projectile-nucleus, respectively;
$$R_{eff}=\frac{R+R_1+R_2}2\,\,, \,\,\,\,
  R_{1(2)eff}=\frac{R_{1(2)}}{R_1+R_2}R\,\,,$$
where  $R_{1(2)}$ is the nucleus radius.

The friction coefficients $\gamma_{R}~
(\gamma_{\theta})$,  {\it i.e.}, the change  in the nucleus-nucleus potential
and reduced mass of relative motion during the interaction time
$t$, are calculated from the estimation of the coupling term
between the relative motion of nuclei and the intrinsic excitation
of nuclei \cite{Adam97}.

The nucleus-nucleus potential  includes Coulomb ($V_C$), nuclear
($V_{nucl}$), and
rotational ($V_{rot}$) potentials:
\begin{equation}
\label{nucnuc}
V(\relR)=V_C(\relR)+V_{nucl}(\relR)+V_{rot}(\relR)+\delta V(R).
\end{equation}
A change $\delta V(R)$ of the nucleus-nucleus potential and
 the  dynamic contribution $\delta \mu(R)$ to the reduced mass
$\mu(R)$
during the interaction time $t$ is taken into account
(see Appendix A of paper \cite{GiarSHE}).

The nucleus-nucleus potential  $V(R)$ depends on the mutual
orientations of the  symmetry axes of the deformed nuclei relative to
$\relR(t)$. Thus, it is possible to consider contributions to the fusion
for different initial orientations of the symmetry axes. The quadrupole ($2^+$)
and octupole ($3^-$) collective excitations in spherical nuclei
are taken into account.   Details of this method of calculation
are presented in   \cite{GiarSHE}.

\subsection{\label{Fusion} Fusion}

The competition between fusion and quasifission is taken into
account by the factor $P_{CN}(E,\ell)$ (fusion factor, hereafter)
which is calculated  in the framework of the statistical
model. This way was firstly used  in \cite{DNSV935}. Validity of
using of the statistical method is righteous due to fact  that at
quasifission  a full relaxation of the relative kinetic
energy and  mass (charge) asymmetry between the two fragments
takes place \cite{Back96}. The
statistical method is acceptable to calculate competition between
complete fusion and quasifission processes due to the fact that
thermal equilibrium is established in the DNS rather fast, for a
few units of $10^{-22}$s:
$$\tau_{ther}=2.6/T_{DNS}^2 \cdot 10^{-22}s.$$
Here $T_{DNS}$ is the effective  temperature of DNS: $T_{DNS}=3.46\sqrt{E^*_{DNS}/A_{tot}}$,
where $E^*_{DNS}=E_{\rm c.m.}-V(R_m)$ is  excitation energy
of DNS; $E_{\rm c.m.}$ is a value of beam
energy  in the system of the center of mass  and $V(R_m)$ is a minimum value of
the nucleus-nucleus potential in the potential well.

Duration of the quasifission is one order of magnitude larger than
$\tau_{ther}$. It is  more than $5\cdot 10^{-21}$ s which was  estimated by the
analysis of experimental data on quasifission reactions
\cite{Velk99,Hinde92,Swiek95}. The fusion time is longer than
quasifission reaction time. The calculation of mass and charge
yields in frame of microscopic model showed that formation of DNS
with the given  mass (charge)  asymmetry changes from $5\cdot
10^{-21}$s to $9\cdot 10^{-20}$s \cite{Obn99,Adamqfis,Butsch91}.
The experimental data on study of fusion-fission and quasifission
reactions induced by $^{48}$Ca and $^{58}$Fe projectiles on $^{232}$Th,
$^{238}$U, $^{248}$Cm and $^{249}$Cf targets \cite{Bogat01,Itkis01} showed
that mass and charge distribution can reach their equilibrium
values even in quasifission reactions. It was observed  that products far
from initial nuclei could be formed not only  at  the fission of a
hot CN but at quasifission of DNS  which lives long enough to reach
mass equilibration.

Experimentally it is difficult to distinguish between fission of
the compound nucleus and quasifission. Only analysis of
correlation between reaction fragment mass and angular
distributions allows us to estimate a ratio between contributions
of quasifission and fusion-fission processes. These theoretical
and experimental results on quasifission justify the use of
statistical approach to estimate competition of the complete
fusion and quasifission processes. 
Calculation of complete fusion in competition with quasifission
can be performed in the framework of statistical methods.
The probability of realizing complete fusion
is related to the ratio of the level densities, depending on the
intrinsic fusion or quasifission barriers, by the expression:
\begin{equation}
\label{Pcn} P_{CN}=\frac{\rho(E^*_{DNS} -
B^*_{fus})}{\rho(E^*_{DNS} - B^*_{fus}) + \rho(E^*_{DNS} -
B_{qf})},
\end{equation}
where $\rho(E^*_{DNS} - B^*_K)$ is the  level   density for the DNS
which is calculated on the quasifission and intrinsic fusion
barriers ($B_K = B_{qf}, B^*_{fus}$)
(all details are in Appendix A of \cite{GiarSHE}).
The final result for the partial fusion cross is obtained by averaging
over the contributions of different mutual orientations of the symmetry
axes of the reacting nuclei.

\subsection{\label{surviv}Survival probability}

The advanced statistical model, described in detail in
\cite{ASM,dar92,Sag98}, allows us to take into account the dynamical
aspects of the fission-evaporation competition during the evolution
of the compound nucleus  along the de-excitation cascade. The
model accounts for the angular momentum and parity
coupling;  it allows for the neutron, proton, and $\alpha$-particle
multiple emission, as well as for the fission channel and full
$\gamma$-cascade in the residual nuclei.

Particular attention is
devoted to the determination of level densities. These are
calculated in the non-adiabatic approach allowing for rotational
and vibrational enhancements. These collective effects are
gradually removed above a certain energy. In the case of
rotational enhancement, this energy is related to the Coriolis
force which couples intrinsic and collective motions.  The used level
densities acquire a dynamic aspect through the dependence of the
Coriolis force and of the rotational enhancement on the nuclear
shape, which is, in turn, obtained from the  classical model of
a rotating liquid drop. Intrinsic level densities are calculated
using the Ignatyuk  approach \cite{Ignat75}, which takes into
account shell structure effects and pairing correlations. Use of
the correct level densities is of fundamental importance for the
present analysis as they determine the phase   space available for
each channel, the very essence that   governs statistical decay.

In the case of evaporation residue production, one should also
carefully consider the low  energy level densities since in this
energy interval most of the evaporation residues
is formed. That is why we use the super-fluid model of the
nucleus \cite{Ignat79} in our calculations, with the standard value
of pairing correction $\Delta = 12/\sqrt{A}$ MeV. The yrast lines are
automatically included in our calculations by the requirement that
the total excitation energy should be higher than the rotational
energy, otherwise the level density  is set to zero.

For the fission barriers, we use the predictions of the rotating
droplet model (angular momentum dependent) as
parameterized by Sierk \cite{Sierk} and allow for angular momentum
and temperature fade-out of the shell corrections \cite{ASM}. This
is expressed by the formula for the actual fission barrier used in
calculations:
\begin{equation}
\label{fissb}
B_{fis}(J,T)=c \ B_{fis}^{m}(J)-h(T) \ q(J) \ \delta W,
\end{equation}
which  includes a dependence on  temperature of the compound nucleus
\[h(T) = \left\{ \begin{array}{ll} 1 \hspace*{3cm} T \leq 1.65 \
\rm{MeV} \\ k\exp{(-mT)}
\hspace*{0.8cm} T > 1.65 \ \rm{MeV},
\end{array}
\right.
\] and \[q(J) = \{ 1 + \exp [(J-J_{1/2})/\Delta J]\}^{-1},
\] where $B_{fis}^{m}(J)$ is the parameterized macroscopic fission
barrier \cite{Sierk} depending on the angular momentum $J$, $\delta W
= \delta W_{sad} - \delta W_{gs} \simeq - \delta W_{gs}$ is the
microscopic (shell) correction to the fission barrier taken from
the tables \cite{Moeller} and the constants for the macroscopic
fission barrier scaling, temperature, and angular momentum
dependencies of the microscopic correction are chosen as
follows: $c = 1.0$, $k = 5.809$, $m = 1.066$ MeV$^{-1}$, $J_{1/2}
= 24 \hbar$ for nuclei with Z $\simeq$ 80--100, $\Delta J = 3
\hbar$. This procedure allows the shell corrections to become dynamical
quantities, also.

Dissipation effects, which delay fission, are treated according to
\cite{GraWeiPLB80,RastSJNP91}. These include Kramers' stationary
limit \cite{KramP40} and an exponential factor applied to Kramers'
fission width to account for the transient time, after which the
statistical regime is reached. The systematics obtained by
Bhattacharya {\it et al}. \cite{BhatPRC96} allows us
to take into account the dependencies of the reduced
dissipation coefficient $\beta_{dis}$ on the incident
energy per nucleon $\epsilon$ and nucleus mass $A$
\begin{equation}
\beta_{dis}(\epsilon,A) = a \epsilon + bA^3,
\end{equation}
 where  $a=0.18$,  $b=0.357\times 10^{-6}$.
$\beta_{dis}$ is the ratio between the
friction coefficient $\gamma$,  which describes
the coupling of the fission degree of freedom to the
intrinsic degrees of freedom.
This ratio characterizes the dissipative and diffusive
motion. For the investigated reactions, the $\beta_{dis}$
values are in   $(6\div 7)\times 10^{21}$ s$^{-1}$ range.

In the present ASM calculations, the target-projectile fusion cross
section was determined by formula (\ref{s_fus}). The survival probability
$W_{sur}$ is defined by the dependence of  fusion cross section
on the initial values of the orbital angular momentum,
since such a spin distribution affects the  fission barrier and the
$\Gamma_n / \Gamma_f$ ratio that determine the evaporation residue
production.

\section{\label{dnsexp}Comparison of calculated
 results and experimental data}

The difference between measured data
on the cross section of evaporation residues for
$^{40}$Ar+$^{176}$Hf \cite{Verm84,Clerc84}, $^{86}$Kr + $^{130}$Xe
\cite{Ogan96} and $^{124}$Sn + $^{92}$Zr \cite{Sahm85}
reactions leading to the heated  $^{216}$Th$^*$, as well as that for
$^{48}$Ca + $^{174}$Yb  \cite{Sagai97} and  $^{86}$Kr + $^{136}$Xe \cite{Ogan96}
reactions leading to the heated  $^{222}$Th$^*$, are explained by the
dependencies of   fusion excitation functions  on the mass asymmetry
and shell structure of colliding nuclei and  by the dependencies of survival
probabilities on  the spin distribution of  the excited  compound nuclei
produced in these reactions.

\subsection{\label{216th}The reactions leading to $^{216}$Th$^*$}

\begin{figure}
\includegraphics[totalheight=13cm]{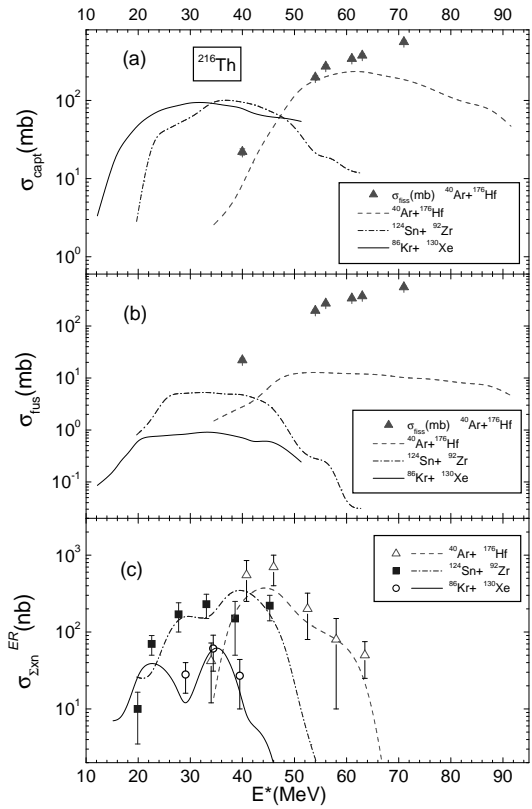}
\vspace*{-2.2cm}
\caption{\label{er216com}Comparison of the calculated capture (a),
fusion (b) and evaporation residue (c) excitation functions  as
well as the measured  excitation functions of evaporation residue
(c)  for $^{40}$Ar+$^{176}$Hf (dashed curve, open up triangles
\cite{Verm84,Clerc84}),
$^{124}$Sn + $^{92}$Zr (dash-dotted curve, solid squares \cite{Sahm85}), and
$^{86}$Kr + $^{130}$Xe (solid curve, open circles \cite{Ogan96}) combinations
leading to the $^{216}$Th$^*$ CN.  The solid triangles  in  (a)  and  (b)  are
 the fission excitation functions obtained from  the  measurements of  the
two symmetric mass fragments for the  $^{40}$Ar+$^{176}$Hf
reaction \cite{Clerc84}.}
\end{figure}

A dependence of the reaction mechanism on the entrance channel
 was studied in experiments with reactions leading to the same
 compound nucleus.
The experimental data reveal  that the maximum value of the ER
cross section for $^{40}$Ar+$^{176}$Hf (I) \cite{Verm84,Clerc84}
is twelve   times larger than for $^{86}$Kr + $^{130}$Xe (II)
\cite{Ogan96} and three  times  larger than for $^{124}$Sn + $^{92}$Zr
(III) \cite{Sahm85} (see Fig.\ref{er216com}). The $^{40}$Ar+$^{176}$Hf
reaction has a larger charge asymmetry
($\eta_{A}=(A_{2}-A_{1})/(A_{1}+A_{2})$) in comparison
with the two others (II,III).
This result agrees with the conclusions of MDM  \cite{Block86}
and DNS models which state that
more asymmetric reactions are favorable for formation of massive
compound nucleus. In MDM, "extra push" energy, which is needed to
transform dinuclear system into compound nucleus, is smaller for
an asymmetric reaction than that for a more symmetric one leading
to the same compound nucleus because
$$Z^{asym}_1 \cdot Z^{asym}_2 < Z^{sym}_1 \cdot Z^{sym}_2$$,
if both of reactions lead to the same compound nucleus
($Z^{asym}_1+ Z^{asym}_2=Z^{sym}_1 + Z^{sym}_2$).
The calculated driving potential   shows that the
barrier $B^*_{fus}$ in the way to fusion (on mass asymmetry
axis) is smaller for asymmetric reaction than that for symmetric one;
the quasifission barrier is larger  for a more asymmetric reaction
(see Table \ref{tabth216}) and as a result  the fusion factor $P_{CN}$  becomes
larger in  this case.
As it is seen  from  Figs.\ref{er216com}a and \ref{er216com}b,
the excitation function  of the  capture and fusion for reaction (I) is
sufficiently higher than that for reactions (II) and (III), because
potential well of entrance channel for the (I) reaction is deeper
than that for  the others.
Therefore, $B^{(I)}_{qf} > B^{(II)}_{qf},\,\,B^{(III)}_{qf}$
(see Table \ref{tabth216}).  The smallness of  $B^*_{fus}$ for the (I) reaction
is connected with the peculiarities of the driving potential (Fig.\ref{dr216}).

The evaporation residue excitation functions  calculated in this paper
using the  method presented in  Section \ref{dnscon}
are in good agreement with the experimental data (see Fig.\ref{er216com}c).
In these calculations, the partial  cross sections of fusion
(\ref{s_fus}) were used.

\begin{table}
\caption{\label{tabth216}
Charge asymmetry, intrinsic fusion ($B^*_{fus}$) and
quasifission ($B_{qf}$) barriers, and the fusion factor ($P_{CN}$)
for the reactions leading to $^{216}$Th$^*$ CN.}
\begin{ruledtabular}
\begin{tabular}{ccccc}
Reactions  &  $\eta_{Z}$ &  $B^*_{fus}$ & $B_{qf}$ & $P_{CN}$ \\
           &             &    (MeV)     &    (MeV) &          \\
\hline\\
 $^{40}$Ar+$^{176}$Hf \quad (I) & 0.63 & 2.31 & 5.62 & 0.121 \\
 $^{86}$Kr + $^{130}$Xe (II) & 0.20 & 12.31& 2.35 & 0.011 \\
 $^{124}$Sn + $^{92}$Zr (III) & 0.15& 9.87  & 1.35 & 0.051 \\
\end{tabular}
\end{ruledtabular}
\end{table}

It is seen from Fig.\ref{er216com}c that the maximum value of the ER
cross section for $^{86}$Kr + $^{130}$Xe is four times smaller than
for $^{124}$Sn + $^{92}$Zr near the same value of $E^*_{CN}$ though the
former reaction  is more asymmetric than latter.
This fact was one of the impact points of  the presented  exploration.
This phenomenon is explained by the driving potential calculated
using binding energies obtained from the mass
table \cite{MassAW95}.  As one can see in top panel of Fig.\ref{dr216},
$B^*_{fus}$ for the $^{86}$Kr + $^{130}$Xe reaction is
larger than the one of the $^{124}$Sn + $^{92}$Zr reaction.
Therefore,  the fusion excitation function is lower for the  former  reaction
than for the latter.  So the observed   difference  between
the excitation functions of   evaporation residues  for  the
$^{86}$Kr + $^{130}$Xe
and  $^{124}$Sn + $^{92}$Zr reactions  is explained
by the difference of  $B^*_{fus}$ calculated for these
reactions using experimental binding energies of fragments.
The values of $B^*_{fus}$ is small  for the region of
the reaction (III) due to shell effects contained in the nuclear
binding energy.

If  the driving potential is calculated using the binding energies
$B_1, B_2$ and $B_{CN}$ obtained in framework of  the
liquid-drop model, then  the intrinsic barriers for  these  two reactions
will be equal  $B^*_{fus}(II) \approx B^*_{fus}(III)$ (see the middle panel
of Fig.\ref{dr216}) and the fusion cross section for the $^{86}$Kr + $^{130}$Xe
reaction will be larger than for the $^{124}$Sn + $^{92}$Zr reaction
due to differences in quasifission barriers.
This is in contradiction with experimental data, which
indicate  that the use of binding energies obtained in the
liquid-drop model is not suitable in such an analysis.
\begin{figure}
\includegraphics[totalheight=11.5cm]{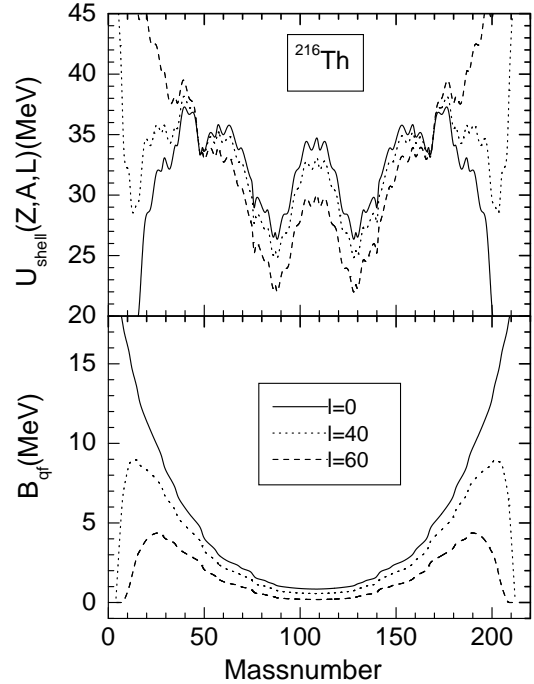}
\vspace*{-2.0cm} \caption{\label{drivL}
The dependence of the driving potential (top panel) and
quasifission barrier (bottom panel) on orbital angular
momentum $\ell (\hbar)$ as a function of
the mass of one of fragments for the reactions leading to $^{216}$Th$^*$ CN.}
\end{figure}

\begin{figure}
\includegraphics[totalheight=12.5cm]{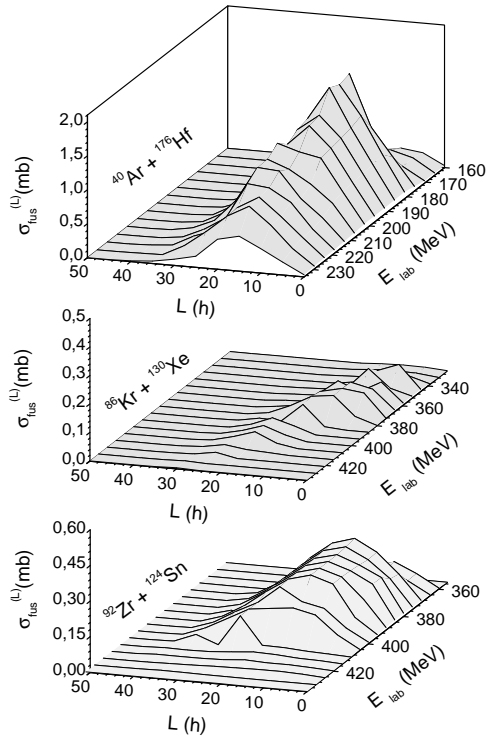}
\vspace*{-2.5cm}
\caption{\label{sp216th}The calculated spin distribution for the
$^{40}$Ar+$^{176}$Hf (top panel),  $^{86}$Kr + $^{130}$Xe (middle),
and $^{124}$Sn + $^{92}$Zr (bottom) reactions at different beam energies
$E_{lab}$.}
\end{figure}

The dependence  of $B^*_{fus}$ on the orbital angular momentum
(Fig.\ref{drivL}) affects  the  partial  cross sections of
 fusion (Fig.\ref{sp216th}).  It is seen from top panel of
Fig.\ref{drivL} that the values of  driving potential for the fragments
of mass  less than $A$=44 increase and the part for masses larger than
$A$=44 decreases by increase of the values of orbital angular momentum.
Consequently, values of $B^*_{fus}$ increases by $\ell$.
But values of quasifission barrier $B_{qf}$ decrease by increasing of the values
of orbital angular momentum (bottom panel of Fig.\ref{drivL}).
As a result the partial fusion cross section
decreases by increase of orbital angular momentum.
This kind  of spin distribution of CN formed in
reaction (I) against the beam energy 
(top panel) has a larger volume in comparison with
reactions (II) (middle panel) and (III) (bottom  panel). But the volume of the
spin distributions of CN corresponding to  reaction (III) is
larger than that for reaction (II). This is  a result of dependence of
the  partial fusion cross sections  $\sigma^{fus}_{\ell}(E)$  (\ref{s_fus})
on the orbital angular momentum of  entrance channel.

As seen  in  middle panel of Fig.\ref{sp216th}, for the
$^{86}$Kr + $^{130}$Xe reaction at lowest values of the orbital
angular momentum, capture becomes impossible for   beam energy
larger than 400 MeV. This is connected  to the small size of the
well in the nucleus-nucleus potential and to the limited value of
 calculated friction coefficient which leads to a gradual
dissipation of relative kinetic  energy \cite{Adam97}.
Therefore,  the dissipation
is not enough to trap colliding nuclei in the potential well.
At the largest values of beam energy the capture is possible only for high
angular momenta (if there is  the potential well for the given $\ell$).
 In this case, the formed DNS can exist in a molecular state, forming
a super-deformed shape, or it undergoes  quasifission because
$B_{fus}^*$ increases with angular momentum of the DNS.
Therefore, the maximum of the calculated spin distributions has a
tendency to move to larger values of angular momentum at beam
energies  well above the Coulomb barrier. It can be seen in the spin
distributions for $^{86}$Kr + $^{130}$Xe  and $^{124}$Sn + $^{92}$Zr
reactions (Fig.\ref{sp216th}).

\begin{figure}[h!]
\includegraphics[totalheight=13.1cm]{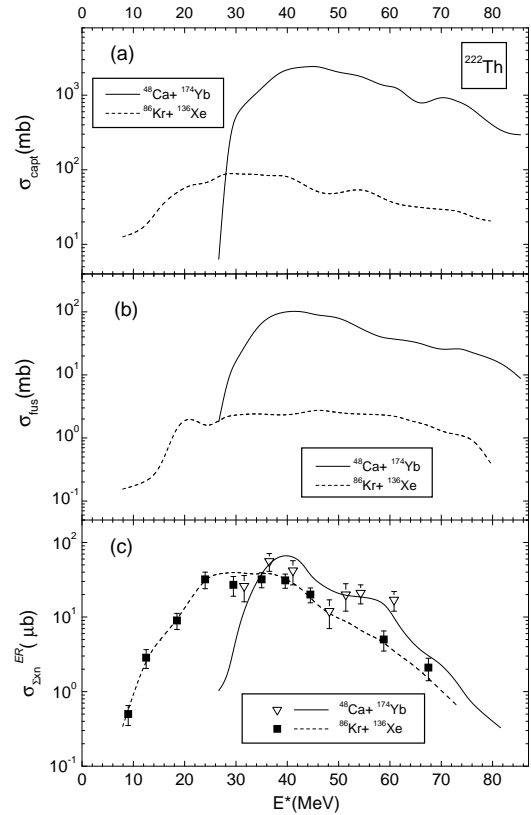}
\vspace*{-1.9cm}
\caption{\label{er222com}Comparison of the calculated capture (a),
fusion (b) and evaporation residue (c) excitation functions as
well as the measured  excitation functions of evaporation residue
(c) for $^{48}$Ca+$^{174}$Yb (solid curve, open down triangles
\cite{Sagai97}) and $^{86}$Kr + $^{136}$Xe (dashed curve, solid
squares  \cite{Ogan96}) reactions leading to $^{222}$Th$^*$.}
\end{figure}

From the analysis of these   (I, II and III)
reactions leading to $^{216}$Th$^*$,  one can conclude that:

-- the influence of the mass asymmetry and
peculiarities of the shell structure on the competition between fusion and
quasifission, and  on the fusion-fission mechanism is strong;

-- the difference between  fusion excitation functions deals  with the values of
$B^*_{fus}$, which depend on the peculiarities of the nuclear
shell structure, and of $B_{qf}$; both $B^*_{fus}$ and $B_{qf}$ depend 
on the entrance channel of reactions under consideration;

-- due to the large difference between the $Q-$values of these three
reactions leading to the  $^{216}$Th$^*$ CN, the centers of their
excitation functions (see Fig.\ref{er216com}) are placed at different
values of excitation energy.

\subsection{\label{ArHffis} Comparison of  capture, fusion and
fission cross sections for the $^{40}$Ar+$^{176}$Hf reaction}

In these reactions under consideration, the evaporation residue
cross sections are several orders of magnitude smaller than the
fission cross sections (Fig.\ref{er216com}). It means  that  fission
cross section is approximately equal to fusion cross section.
Comparison of the calculated  fusion excitation function
and the measured  fission excitation function is intriguing
when discussing the mechanism of fusion-fission reactions.
This  has  been done for the $^{40}$Ar+$^{176}$Hf  reaction
(Fig.\ref{er216com}b). In \cite{Clerc84}, the fission excitation
function  was obtained  from the detection of reaction
products of symmetric masses.
It should be stressed that those products  could be formed
not only  at  the fission of a hot CN but at
quasifission of DNS which lives long enough to
reach mass equilibration.
Experimentally it is difficult to distinguish between fission of
the compound nucleus and quasifission. In \cite{GiarArAg}, the
calculations showed that the contribution of quasifission is increased
with beam energy above the fusion barrier.
For this reason, the measured fission data in the
$^{40}$Ar+$^{176}$Hf reaction \cite{Clerc84} are closer to the
calculated excitation function of  capture  (Fig.\ref{er216com}a)
that is a sum of fusion and quasifission cross sections.

Therefore, the fact that the measured fission cross section  is
higher than the calculated fusion cross section (Fig.\ref{er216com}b) could be
explained by the sizeable contribution of quasifission products
to the measured fission data \cite{Clerc84}.
Appearance of difference between the measured fission
 and theoretical capture cross sections at energies higher than
$E^*=55$ MeV means that events of capture accompanied by
the pre-equilibrium emission of neutrons, protons and $\alpha$-
particles from fragments were not taken into account in the
model under consideration.

\subsection{\label{222th}The reactions leading to  $^{222}$Th$^*$}

The maximum of the experimental excitation functions of evaporation
residues for $^{48}$Ca + $^{174}$Yb (IV) \cite{Sagai97} is
higher than that for $^{86}$Kr + $^{136}$Xe (V) \cite{Ogan96}
(Fig.\ref{er222com}c).
This fact can be explained by the large fusion cross sections at
excitation energies $E^*$  higher than  24 MeV  (Fig.\ref{er222com}b).
These reactions lead to the $^{222}$Th$^*$ CN.
Excitation functions of capture and fusion
for the $^{48}$Ca + $^{174}$Yb reaction are more than one order
of magnitude higher  than for the  $^{86}$Kr + $^{136}$Xe reaction.

In Table \ref{tabth222}, we report the values of the charge
asymmetry, intrinsic fusion barrier
and quasifission barrier  for such two reactions leading to the
$^{222}$Th$^*$ CN.
At excitation energies $E^*$  of the $^{222}$Th$^*$ CN lower
than about 30 MeV, the excitation functions of capture,
fusion and evaporation residue go down for the $^{48}$Ca +
$^{174}$Yb reaction. This energy  corresponds to the  Coulomb
barrier.     Because the $Q_{gg}$-value for this reaction
(-118.35 MeV) is not as low as for the $^{86}$Kr + $^{136}$Xe
(-186.88 MeV) reaction, the subbarrier region of fusion for the
$^{48}$Ca + $^{174}$Yb reaction is placed at $E^*<$ 30 MeV.

\begin{figure}
\includegraphics[totalheight=9cm]{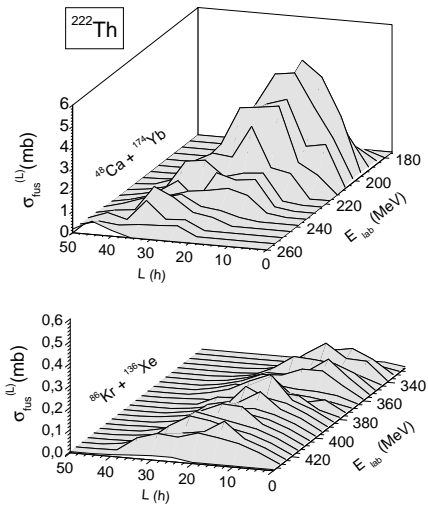}
\vspace*{-0.75cm}
\caption{\label{sp222th}Spin distribution for
$^{48}$Ca+$^{174}$Yb (top panel) and  $^{86}$Kr + $^{136}$Xe
(bottom panel) reactions at different beam energies $E_{lab}$.}
\end{figure}

By considering the spin distributions  of the two  reactions
leading to $^{222}$Th$^*$ CN,
we find a higher contribution to the fission process for the
$^{48}$Ca + $^{174}$Yb reaction caused by the spin distribution
peaked at higher $J$ spin values. As one can see, the spin
distribution of CN formed in the 
$^{48}$Ca + $^{174}$Yb reaction against the beam energy
(Fig.\ref{sp222th}, top panel) has a larger volume than that of
the $^{86}$Kr + $^{136}$Xe reaction (bottom panel).

From calculations of survival probability, we find that in the
range of excitation energy between 40 and 60 MeV of \,
$^{222}$Th$^*$, and for various steps of the de-excitation cascade, the
$\Gamma_n/\Gamma_f$ ratio values for the $^{48}$Ca + $^{174}$Yb reaction
are much  lower than those for the $^{86}$Kr + $^{136}$Xe reaction.
The compound nucleus $^{222}$Th$^*$ formed in these reactions at the
same excitation energy $E^*$ has different spin distributions which are
caused by the dynamical effects in the entrance channel of the two reactions.
Due to dependence of fissility on spin distribution in ASM calculation,
the evaporation residue have different cross sections.
The ratio $(\Gamma_n/\Gamma_f)_{(V)} / (\Gamma_n/\Gamma_f)_{(IV)}$
of the $\Gamma_n/\Gamma_f$  values for the $^{48}$Ca + $^{174}$Yb and
$^{86}$Kr + $^{136}$Xe reactions at each step of de-excitation
cascade ranges between 2.5 and $6.6\times 10^4$.
Even if in such an energy range the fusion cross section for the
$^{48}$Ca + $^{174}$Yb reaction is about 1-2 orders of magnitude higher than
that  for the $^{86}$Kr + $^{136}$Xe reaction, the survival probability
$W_{sur}(\Gamma_n/\Gamma_f)$ makes the values of the ER cross section
 for  $^{48}$Ca + $^{174}$Yb reaction only a
factor 2-4 times higher than the ER values of $^{86}$Kr + $^{136}$Xe.

\begin{table}
\caption{\label{tabth222}
Charge asymmetry, intrinsic fusion ($B^*_{fus}$) and
quasifission ($B_{qf}$) barriers and the fusion factor ($P_{CN}$)
 for the reactions leading to $^{222}$Th$^*$ CN.}
\begin{ruledtabular}
\begin{tabular}{ccccc}
Reactions  &  $\eta_{Z}$ &  $B^*_{fus}$ & $B_{qf}$ & $P_{CN}$ \\
           &             &    (MeV)     &    (MeV) &          \\
\hline\\
 $^{48}$Ca+$^{174}$Yb (IV) & 0.56 & 3.20 & 5.37 & 0.065 \\
 $^{86}$Kr + $^{136}$Xe (V) & 0.20 & 7.52 & 4.05 & 0.027 \\
\end{tabular}
\end{ruledtabular}
\end{table}

The analysis of the $^{48}$Ca + $^{174}$Yb and the $^{86}$Kr + $^{136}$Xe reactions
leading to $^{222}$Th$^*$ CN shows that:

--  influence of the mass asymmetry and
peculiarities of the shell structure on the competition between fusion and
quasifission  mechanism is strong. Nevertheless  the  comparison of the measured data
on the cross section of evaporation residues does not  reflect the role of
mass asymmetry of entrance channel. The large difference between  the fusion cross sections
 was compensated by the different fissility
of nuclei formed in these reactions at various step of de-excitation cascade;

-- the difference between  fusion excitation functions deals  with the values of
$B^*_{fus}$ and  difference between survival probabilities is
connected with the dependence of fusion cross sections on
the orbital angular momentum in the entrance channel of reactions
under consideration.

\section{\label{130136xe}Comparison of reactions induced by  $^{86}$Kr
on the $^{130}$Xe and $^{136}$Xe targets}

\begin{figure}
\includegraphics[totalheight=13cm]{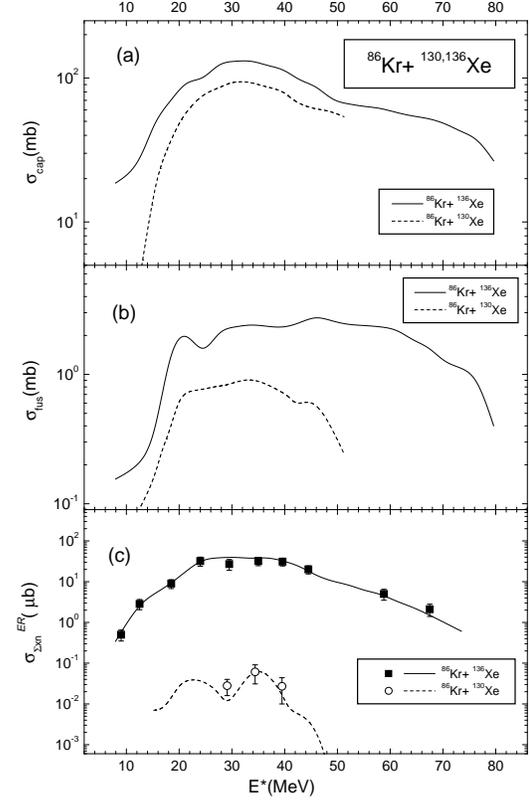}
\vspace*{-2.2cm}
\caption{\label{er30xe36}Comparison of the calculated capture (a),
fusion (b) and evaporation residue (c) excitation functions as
well as the measured  excitation functions of evaporation residue
(c) for the $^{86}$Kr + $^{136}$Xe (solid curve, solid squares \cite{Ogan96})
and  $^{86}$Kr + $^{130}$Xe (dashed curve, open circles \cite{Ogan96})
reactions.}
\end{figure}

Another  interesting phenomenon  which was  observed
in the comparison of the experimental data for reactions induced by
the $^{86}$Kr projectile on the $^{130}$Xe and $^{136}$Xe targets
is that  the ER  cross section in $^{86}$Kr + $^{130}$Xe (II)
 was about 500 times smaller than that in  $^{86}$Kr + $^{136}$Xe (V)
 (Fig.\ref{er30xe36}c).
The  experimental and theoretical excitation functions presented
in  Fig.\ref{er30xe36}c are  the sum of the evaporation residues
along the de-excitation cascade for the neutron emission from
$^{216}$Th$^*$ and  $^{222}$Th$^*$ CN formed in these  reactions, respectively. 
It is clear that these
differences  are caused by the excess number of neutrons in the
$^{136}$Xe target  in comparison with the $^{130}$Xe one.

\begin{figure}
\includegraphics[totalheight=12cm]{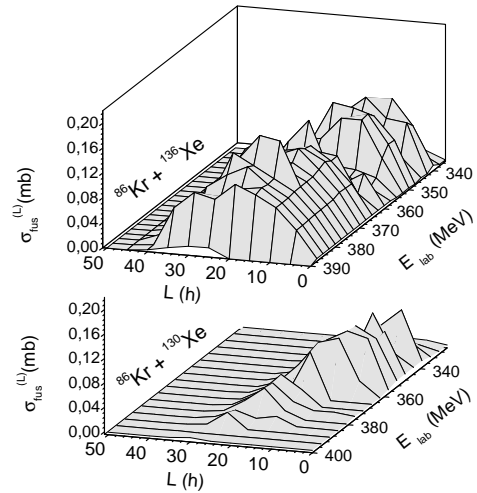}
\vspace*{-4cm}
\caption{\label{130sp136}Spin distribution for the
$^{86}$Kr+$^{136}$Xe (top panel) and  $^{86}$Kr + $^{130}$Xe
(bottom panel) reactions at different beam energies  $E_{lab}$.}
\end{figure}

As a result  we obtain  differences in  two characteristics of
the fusion-fission mechanism:

--   the fusion cross section calculated using the  model based on
DNS concept   \cite{GiarSHE,GiarArAg}  for the reaction $^{86}$Kr + $^{130}$Xe
is much smaller than the one for $^{86}$Kr + $^{136}$Xe (Fig.\ref{er30xe36}b).
Therefore, the volume of spin distribution of CN formed 
in the former reaction against the beam energy is
smaller than that  for the last reaction (Fig.\ref{130sp136}). This is because
for the $^{86}$Kr + $^{136}$Xe reaction  the intrinsic fusion barrier  is smaller
and the quasifission barrier is larger   than those  for the 
$^{86}$Kr + $^{130}$Xe reaction (see Tables \ref{tabth216} and \ref{tabth222});

--  the survival probability ($W_{sur}$) decreases along the steps of the
$^{216}$Th* de-excitation cascade, while $W_{sur}$ increases along
the steps of the $^{222}$Th* cascade.
This is because the shell corrections, in average, decrease for the
intermediate excited nuclei after $1n, 2n \ldots xn$ emissions from
$^{216}$Th$^*$, whereas the shell corrections of the
excited nuclei increase after the analogous neutron emission from
$^{222}$Th$^*$. Notice concerning  to the excitation energy
of the initial compound nucleus,
the ER production is more sensitive to the last step of the
de-excitation cascade. Moreover, we have to note that at each step
(1$n$, 2$n$, 3$n$ $\ldots$ x$n$) of the de-excitation cascade of the
initial  compound nucleus, the neutron separation
energy $S_n$ at each step of the $^{222}$Th* decay chain
is about 1-2 MeV lower than that at the analogous step
of the neutron emission from $^{216}$Th*.
By comparing the $\Gamma_n/\Gamma_f$ values at
each step of the cascade of $^{222}$Th* and $^{216}$Th*, at the same
excitation energy of the CN,  we find  that
$(\Gamma_n/\Gamma_f)_{^{222}Th*}$ are always much larger
than $(\Gamma_n/\Gamma_f)_{^{216}Th*}$.
In particularly, the  $(\Gamma_n/\Gamma_f)_{^{222}Th*} /
(\Gamma_n/\Gamma_f)_{^{216}Th*}$  ratio ranges  between
$2.7\times 10^4$ and $9.1\times 10^5$ for the 4$n$-channel.
Since large $\Gamma_n/\Gamma_f$ values correspond to a large evaporation
residue cross section, the excess number of neutrons increases
the survival probability in the $^{86}$Kr+$^{136}$Xe reaction
in comparison with the $^{86}$Kr+$^{130}$Xe one.

\section{\label{concl}Conclusions}

The role of the entrance channel in fusion-fission reactions was
studied  intending  to account for the difference between
the experimental data  for  the $^{40}$Ar + $^{176}$Hf \cite{Verm84,Clerc84},
$^{86}$Kr + $^{130}$Xe \cite{Ogan96} and $^{124}$Sn + $^{92}$Zr \cite{Sahm85}
reactions leading to the $^{216}$Th$^*$ compound nucleus, 
and the $^{48}$Ca + $^{174}$Yb \cite{Sagai97}, $^{86}$Kr + $^{136}$Xe \cite{Ogan96}
 reactions leading to the $^{222}$Th$^*$ compound nucleus.
The results of calculations in the framework of the DNS concept \cite{GiarSHE}
for the fusion cross sections, 
 and advanced statistical model \cite{ASM,dar92,Sag98} for the total
evaporation residue cross sections  have been compared with the  
measured experimental data for these reactions. 
From the analysis of the  experimental  data  four phenomena were studied:

i)  Among  reactions leading to $^{216}$Th$^*$,
$^{40}$Ar + $^{176}$Hf  has  more larger evaporation residues
in comparison with two others: $^{86}$Kr + $^{130}$Xe  and
$^{124}$Sn + $^{92}$Zr. This result affirms conclusions of  macroscopic
dynamical and DNS models  which state that more
asymmetric reactions are favorable to formation of massive compound nucleus.
In MDM, "extra push" energy, which is needed to transform dinuclear system
into compound nucleus, is smaller for
asymmetric reaction than for more  symmetric one.
The "extra push" energy in MDM  \cite{Block86} and the 
intrinsic fusion barrier $B^*_{fus}$ in DNS concept  \cite{DNSV935}, 
(both of which  are a hindrance to fusion)
are  smaller for an asymmetric reaction than for more  symmetric one leading
to the same compound nucleus  (top panel, Fig.\ref{dr216}).

ii)  One of the unexpected  phenomenon is that the measured maximum
value of the ER cross section for  $^{86}$Kr + $^{130}$Xe (II)  is four times
smaller than that for $^{124}$Sn + $^{92}$Zr (III) , nearly at the same
$E^*$ value. This result  is in opposite tendency to the
conclusions of MDM  and  DNS models. The observed   difference  between
the excitation functions of   evaporation residues  for  the
$^{86}$Kr + $^{130}$Xe and  $^{124}$Sn + $^{92}$Zr reactions 
is explained by the difference of  $B^*_{fus}$ calculated for these
reactions using experimental binding energies of fragments
\cite{MassAW95}. 
As one can see in top panel of Fig.\ref{dr216},
$B^*_{fus}$ for the $^{86}$Kr + $^{130}$Xe reaction is
larger than that of the $^{124}$Sn + $^{92}$Zr reaction.
Therefore,  the fusion excitation function is lower for the  former  reaction
than for the latter.  
The  calculated partial cross sections of  fusion 
depend on these intrinsic fusion and quasifission
barriers.  The volume under surface $\sigma_{\ell}^{fus}(E)$
calculated against beam energy for the  $^{86}$Kr + $^{130}$Xe 
reaction  is smaller than that for the  $^{124}$Sn + $^{92}$Zr 
reaction (Fig.\ref{sp216th}).
This created the necessary prerequisites  to obtain larger 
cross sections of the evaporation residue
for the latter reaction in comparison with former one. 
The calculated results are in good agreement with the experimental data.

If driving potential is  calculated using binding energies of
liquid-drop model (see middle panel of Fig.\ref{dr216}),
the intrinsic fusion barriers $B^*_{fus}$ for the $^{86}$Kr + $^{130}$Xe
and $^{124}$Sn + $^{92}$Zr reactions are almost the same
and the fusion cross section for the former reaction will be larger
than for the latter due to differences in quasifission barriers.
That would  be in contradiction to the experimental data.

iii)   The maximum of the experimental excitation functions of evaporation
residues for $^{48}$Ca + $^{174}$Yb (IV) \cite{Sagai97} is
higher than that for $^{86}$Kr + $^{136}$Xe (V) \cite{Ogan96}
(Fig.\ref{er222com}c).
This fact can be explained by the large fusion cross section at
excitation energies  $E^*>25$ MeV  (Fig.\ref{er222com}b).
These reactions lead to the $^{222}$Th$^*$ CN.
Excitation functions of capture and fusion
for the $^{48}$Ca + $^{174}$Yb reaction are more than one order
of magnitude higher  than for the  $^{86}$Kr + $^{136}$Xe reaction.
But  due to strong dependence of the various steps of the 
de-excitation cascade on the spin distribution of hot and rotated 
compound nuclei: the values of the
$\Gamma_n/\Gamma_f$ ratio for the $^{48}$Ca + $^{174}$Yb reaction
are much lower than those for the $^{86}$Kr + $^{136}$Xe reaction.
The two  different entrance channels
does not produce the same evaporation residue cross section due to
a different fissility of the compound nucleus $^{222}$Th$^*$. Different 
spin distributions $\sigma_{fus}^{(L)}$  are caused by the dynamical
effects in the entrance channel of the two very different reactions.
Due to dependence of fissility on spin distribution in ASM calculation,
the evaporation residues have different cross sections.

The  comparison of the measured data
on the cross section of evaporation residues does not  reflect the role of
mass asymmetry of entrance channel  (Fig.\ref{er222com}c).
The large difference between  the fusion cross sections
was compensated by the different fissility
of nuclei formed in these reactions at various steps of de-excitation cascade.

iv)  Another  interesting phenomenon  which was  observed
in the comparison of the experimental data for reactions induced by
the $^{86}$Kr projectile on the $^{130}$Xe and $^{136}$Xe targets
is that  the ER  cross section in $^{86}$Kr + $^{130}$Xe (II) was about
500 times smaller than that in $^{86}$Kr + $^{136}$Xe (IV) (Fig.\ref{er30xe36}c).
The  experimental and theoretical excitation functions presented
in  Fig.\ref{er30xe36}c are  the sum of the evaporation residues
along the de-excitation cascade for the $^{86}$Kr + $^{130}$Xe and
$^{86}$Kr + $^{136}$Xe reactions.
These differences  are caused by the excess number of neutrons in the
$^{136}$Xe target  in comparison with  $^{130}$Xe one.
Analysing  the mechanism  of these reactions, we conclude that
due to smallness  of intrinsic fusion barrier and  largeness of
quasifission barrier,   capture and fusion cross sections of reaction with  the
$^{136}$Xe target is larger than that the $^{130}$Xe one
(see Tables \ref{tabth216} and \ref{tabth222}).  The excess number of neutrons
increases the survival probability  in the $^{86}$Kr + $^{136}$Xe
reaction due to increase of  values  of the $\Gamma_n/\Gamma_f$  ratio
in comparison with $^{86}$Kr + $^{130}$Xe reaction.

To analyse fusion-fission process,  the fission cross sections
presented in \cite{Clerc84} for the $^{40}$Ar+$^{176}$Hf
reaction  were compared with the calculated fusion and capture cross sections.
The calculated capture cross sections 
are in agreement with the measured fission data  \cite{Clerc84}
up to excitation energies  $E^*$ of about 54 MeV (Fig.\ref{er216com}a).
Due to sizeable contribution of quasifission products to
the measured fission data \cite{Clerc84}, the last are larger
than the calculated fusion cross section  (Fig.\ref{er216com}b).
The deviation of the calculated capture cross sections from the
measured fission data at $E^*>54$ MeV is connected to the fact
that the pre-equilibrium emission of neutrons, protons and $\alpha$-
particles from fragments were not taken into account in the
model under consideration.

In summary, the difference between measured data
on the cross section of evaporation residues for  reactions
leading to the same compound nuclei can be explained by  the
difference in the excitation functions of fusion or survival
probability of the excited  compound nucleus. Decrease of  fusion
cross sections is connected by increase of events coming from
the quasifission process. Competition between complete fusion and
quasifission depends on the dynamics of the entrance
channel and the  nuclear shell structure for colliding nuclei.
The formation of a compound nucleus at low excitation energy does not
ensure the production of evaporation residues with a larger cross section.

\begin{acknowledgments}
 This work was performed partially under the financial support of the RFBR
(Grant No. 99-02-16447) and INTAS (Grant No. 991-1344). We are grateful to
Profs. R.V. Jolos, V.V. Volkov and  W. Scheid; Drs. G.G. Adamian and N.V.
Antonenko for the helpful discussions. One of the authors (A.K.N.) thanks the
Heisenberg-Landau Program for support while staying at the GSI, and RFBR
(Grant No. 01-02-16033) for financial support. Authors (A.I.M and A.K.N.)
are grateful to the STCU  Uzb-45, Uzbekistan State Scientific-Technical
Committee (Grant No. 7/2000) and Fund of Uzbek Academy of Science
for Support of Basic Research (No. 45-00)  for partial support.
 A.K.N. would like to express his gratitude for the warm hospitality
during his stay at GSI, Giessen Justus-Liebig University (Germany),
and Universit{\`a} di Messina (Italy).
R.N.S. and A.K.N. are also grateful to the Fondazione Bonino-Pulejo
(FBP) of Messina for the support received in the collaboration with
the Messina group.
\end{acknowledgments}

\end{document}